\documentclass[11pt,twoside]{article}
\usepackage{fancyhead}
\usepackage[]{algorithm2e}
\usepackage{amssymb,amsmath}

\newcounter{proposition}

\newcommand{\nothing}[1]{}

\newcommand{\beq}[1]{\begin{equation}\label{#1}}
	\newcommand{\eeq}{\end{equation}}

\newcommand{\bmu}[1]{\begin{multline}\label{#1}}
	\newcommand{\emu}{\end{multline}}

\newcommand{\eq}{\triangleq}

\renewcommand{\emptyset}{\varnothing}

\renewcommand{\varlimsup}{\mathop{\overline{\lim}}\limits}

\newcommand{\x}{{\textbf{\textit{x}}}}

\newcommand{\e}{{\bf e}}

\newcommand{\F}{\mathcal{F}}

\renewcommand{\S}{{\cal S}}

\renewcommand{\L}{{\bf L}}

\renewcommand{\l}{\ell}

\renewcommand{\S}{{\mathcal{S}}}
\renewcommand{\L}{{\mathcal{L}}}

\newcommand{\s}{{ {s}}}

\renewcommand{\chi}{\upsilon}
\renewcommand{\l}{{ {L}}}

\renewcommand{\[}{\left[}

\renewcommand{\le}{\leqslant}
\renewcommand{\ge}{\geqslant}

\renewcommand{\l}{\ell}
\renewcommand{\sectionmark}[1]{}
\renewcommand{\subsectionmark}[1]{}
\begin{document}

\vspace*{5mm}

\noindent
\textbf{\LARGE Adaptive Learning a Hidden Hypergraph
\footnote{The research is supported in part by the Russian Foundation for Basic Research under
Grant No. 16-01-00440.}}
\thispagestyle{fancyplain} \setlength\partopsep {0pt} \flushbottom
\date{}

\vspace*{5mm}
\noindent
\textsc{A. G. D'yachkov} \hfill \texttt{agd-msu@yandex.ru} \\
{\small Lomonosov Moscow State University, Moscow, Russia} \\[3pt]
\textsc{I.V. Vorobyev} \hfill \texttt{vorobyev.i.v@yandex.ru} \\
{\small Lomonosov Moscow State University, Moscow, Russia} \\[3pt]
\textsc{N.A. Polyanskii} \hfill \texttt{nikitapolyansky@gmail.com} \\
{\small Lomonosov Moscow State University, Moscow, Russia} \\[3pt]
\textsc{V.Yu. Shchukin} \hfill \texttt{vpike@mail.ru} \\
{\small Lomonosov Moscow State University, Moscow, Russia} \\[3pt]

\medskip

\begin{center}
\parbox{11,8cm}{\footnotesize
\textbf{Abstract.} 
Learning a hidden hypergraph is a natural generalization of the classical group testing problem that consists in detecting unknown hypergraph $H_{un}=H(V,E)$ by carrying out edge-detecting tests.
In the given paper we focus our attention only on a specific family $\F(t,s,\l)$  of localized hypergraphs for which  the total number of vertices $|V| = t$, the number of edges $|E|\le s$, $s\ll t$, and the cardinality of any edge $|e|\le\l$, $\l\ll t$. Our goal is to identify all edges of $H_{un}\in \F(t,s,\l)$ by
using the minimal number of tests. We provide an adaptive algorithm that matches the information theory bound, i.e., the total number of tests of the algorithm in the worst case is at most $s\l\log_2 t(1+o(1))$.}
\end{center}

\baselineskip=0.9\normalbaselineskip

\section{Introduction}
\lhead{}
\rhead{}
\chead[\fancyplain{}{\small\sl\leftmark}]{\fancyplain{}{\small\sl\leftmark}}
\cfoot{}
\markboth{\hspace{-0.2cm}Fifteenth International Workshop on Algebraic and Combinatorial Coding Theory\\ June 18-24, 2016, Albena, Bulgaria \hfill pp. 139--144}{}
\setcounter{page}{139}
Before we introduce the problem, let us recall some definitions and notations. 

Let
$|A|$ denote the size of a set $A$, and $[N]\eq\{1,2,\dots,N\}$ - the set of integers from~$1$ to~$N$. A \textit{hypergraph} is  a pair $H=H(V,E)$ such that $E\subset 2^{V}\setminus \varnothing$, where $V$ is the set of vertices and $E=\{\e_1,\dots \e_s\}$ is a set of edges. A vertex $v\in V$ is called \textit{active}, if there exists at least one edge $\e\in E$ such that $v\in e$. A set $S\subset V$ is called an \textit{independent} set of $H$ if it contains no entire edge of $H$. We denote by $\dim(H)$ the  cardinality of the largest edge, i.e. $\dim(H)=\max\limits_{\e\in E}|\e|$.
\subsection{Statement of the problem}
The problem of  learning a hidden hypergraph is described as follows. Suppose there is an unknown (hidden) hypergraph $H_{un} = H(V,E)$ whose edges are not known to us, but we know that the unknown hypergraph $H_{un}$ belongs to some family $\F$ of hypergraphs that have a specific structure (e.g, $\F$ consists of all Hamiltonian cycles on $V$).  Our goal is to identify all edges of $E$ by carrying out the minimal number $N$ of \textit{edge-detecting queries} $Q(S)$, where $S\subseteq V$:  $Q(S)$ = 0 if $S$ is independent of $H_{un}$, and  $Q(S)$ = 1 otherwise.

In the given paper we focus our attention only on the family of localized hypergraphs. We consider the family $\F(t,s,\l)$, that consists of all hypergraphs $H(V,E)$  such that  $\dim(H)\leqslant \l$ and $|E|\le s$. Suppose we know that the hypergraph $H_{un}$ belongs to the family $\F(t,s,\l)$.  An algorithm is said to be \textit{$\F(t,s,\l)$-searching} algorithm if it finds $H_{un}$, i.e. there exists only one hypergraph from $\F(t,s,\l)$ that fits all answers to the queries.

One of the most important aspects  of any searching strategy is its adaptiveness. An algorithm is \textit{non-adaptive} if all queries are carried out in parallel. An algorithm is \textit{adaptive} if the later queries may depend on the answers to earlier queries. 

By $N^{na}(t,s,\l)$ ($N^{a}(t,s,\l)$) denote the minimal number of queries in a $\F(t,s,\l)$-searching non-adaptive (adaptive) algorithm. Introduce the\textit{ asymptotic rate} for optimal $\F(t,s,\l)$-searching algorithms:
$$
R^{na}(s,\l) \eq \varlimsup\limits_{t\to\infty}\frac{\log_2 t}{N^{na}(t,s,\l)}, \qquad R^{a}(s,\l) \eq \varlimsup\limits_{t\to\infty}\frac{\log_2 t}{N^{a}(t,s,\l)}.
$$

The given paper is organized as follows. In Sect. \ref{prev}, we discuss previously known results and remind the concept of cover-free codes which is close to the subject.  In Sect. \ref{main}, we present the main result of the paper and provide the deterministic adaptive algorithm that matches the information theory bound.

\section{Previous Results}\label{prev}
\headrulewidth 0pt
\lhead[\fancyplain{}{\thepage}]{\fancyplain{}{\rightmark}}
\rhead[\fancyplain{}{\leftmark}]{\fancyplain{}{\thepage}}
\lfoot{}
\rfoot{}
\chead{}
\cfoot{}
\markboth{\textsl{ACCT2016}}{\textsl{D'yachkov, Vorobyev, Polyanskii, Shchukin}}

 For the particular case $\l=1$, the above definitions were already introduced to describe the model called {\em designing screening experiments}. It is a classical group testing problem. We refer the reader to the monograph  \cite{DH} for a survey on group testing and its applications.  It is quite clear (e.g., see \cite{DH}) that a $\F(t,s,1)$-searching adaptive algorithm can achieve the information theory bound, i.e. $N(t,s,1) = s\log_2 t (1+o(1))$ as $t\to\infty$. Therefore, $R^{a}(s,1) = 1/s$. 
 
If $\l =2$, then we deal with learning a hidden graph. One important application area for such problem is bioinformatics \cite{bg05}, more specifically, chemical reactions and genome sequencing.   Alon et al. \cite{ab04}, and Alon and Asodi \cite{aa05} give lower and upper bounds on the minimal number of tests for
non-adaptive searching algorithms for certain families of graphs, such as stars, cliques, matchings. In \cite{bg05}, Boevel et al. study the problem of reconstructing a Hamiltonian cycle. In~\cite{a08}, Angluin et al. give a suboptimal $\F(t,s,2)$-searching adaptive algorithm. More precisely, they prove $R^{a}(s,2)\ge 1/(12s)$.

For the general case of parameters $s$ and $\l$, Abasi et al. have recently provided  \cite{AB14} a suboptimal $\F(t,s,\l)$-searching adaptive algorithm. In particular, from their proofs it follows $R^{a}(s,\l) \ge 1/(2s\l)$. This bound differs up to the constant factor from the information theory upper bound $R^{a}(s,\l) \le 1/(s\l)$.
\subsection{Cover-Free Codes}
A binary $N\times t$-matrix
\beq{X}
X=\|x_i(j)\|, \quad x_i(j)=0,1,\;i\in[N],\; j\in[t]
\eeq
 is called a {\em  code of length $N$  and size $t$}. By $\x_i$ and $\x(j)$ we denote the $i$-th row and the $j$-th column of the code $X$, respectively.

Before we give the well-known definition of cover-free codes, note that any $\F(t,s,\l)$-searching non-adaptive algorithm consisting of $N$ queries can be represented by a binary $N\times t$ matrix $X$ such that each test corresponds to the row, and each vertex stands for the column. We put $x_i(j)=1$ if the $j$-th vertex is included to the $i$-th test; otherwise, $x_i(j)=0$.

\textbf{Definition 1.}\quad
A code $X$ is called a {\em cover-free $(s,\ell)$-code}
(briefly, {\em CF $(s,\ell)$-code}) if for any two non-intersecting sets
$\S,\,\L\subset[t]$, $|\S|=\s$, $|\L|=\ell$, $\S\cap\L=\varnothing$,
there exists a row $\x_i$, $i\in [N]$, for which
\beq{property}
\begin{aligned}
&x_i(j)=0 \; \text{for any}\;  j\in\S,\quad
&x_i(k)=1\; \text{for any}\; k\in\L.
\end{aligned}
\eeq
Taking into account the evident symmetry over  $s$  and $\l$, we introduce
$N_{cf}(t,s,\ell)=N_{cf}(t,\ell,s)$ - the minimal length of
CF $(s,\ell)$-codes of size  $t$ and define
the {\em rate} of  CF $(s,\ell)$-codes:
\beq{Rsl}
R_{cf}(s,\ell)=R_{cf}(\ell,s)\eq \varlimsup_{t\to\infty}\frac{\log_2 t}{N_{cf}(t,s,\l)}.
\eeq
In  \cite{dv02}, Dyachkov et al. show that any CF $(s,\l)$-code represents a $\F(t,s,\l)$-searching non-adaptive algorithm, while any $\F(t,s,\l)$-searching non-adaptive algorithm  corresponds to both a CF $(s,\l-1)$-code and CF $(s-1,\l)$-code. The best presently known  upper and lower bounds on the rate  $R(s,\l)$ of CF $(s,\l)$-codes were presented in~\cite{d14}.
If $\l\ge1$ is fixed and $s\to\infty$, then these bounds lead to the following asymptotic equality:
\beq{limits}
\frac{(\l+1)^{\l+1}}{2e^{\l-1}}\frac{\log_2 s}{s^{\l+1}}(1+o(1))\ge R^{na}(s,\l) \simeq R_{cf}(s,\l) \ge \frac{\l^\l}{e^{\l}}\frac{\log_2 e}{s^{\l+1}}(1+o(1)).
\eeq
\section{New Result}\label{main}
By a counting argument, the lower bound is true.

\textbf{Theorem 1.}\quad \textit{Any $\F(t,s,\l)$-searching algorithm has at least $s\l \log_2{t}(1+o(1))$ edge-detecting queries. In other words, the rate $R(s,\l)\le 1/(s\l)$.}

The key result of this paper is given as follows. 

 \textbf{Theorem 2.}\quad \textit{There exists an adaptive $\F(t,s,\l)$-searching algorithm which has at most $s\l \log_2{t}(1+o(1))$ edge-detecting queries. In other words, the rate $R^{a}(s,\l) = 1/(s\l)$.}
 
\textbf{Proof of Theorem 2.}

We present the full description of $\F(t,s,\l)$-searching algorithm by Alg. \ref{AlgoMain}, and this algorithm is based on Alg. \ref{AlgoSearchEdges}, \ref{AlgoSearchVertex} and \ref{AlgoSearchQuery}. Notice that Alg. \ref{AlgoSearchVertex} is a variation of the binary vertex search. Also one can check that at each step of the algorithm, set $S'$ contains at least one new active vertex. Alg. \ref{AlgoSearchEdges} and \ref{AlgoSearchQuery} represent an exhaustive  search of edges and an exhaustive query search, respectively.

Now we upper bound the number of tests of Alg. \ref{AlgoMain} in the worst scenario. Let $|V|=t$. It is easy to check that Alg. \ref{AlgoSearchVertex} makes use of at most $\lceil \log_2 |S|\rceil\le \lceil \log_2 t\rceil$ tests. One can see that the number of active vertices of the hidden hypergraph $H_{un}\in\F(t,s,\l)$ is at most $s\l$. Alg. \ref{AlgoSearchEdges} uses at most $F_{1}(s,\l)$ tests, while Alg. \ref{AlgoSearchQuery} uses at most $F_{2}(s,\l)$ tests, where the functions $F_{1}$  and $F_{2}$ do not depend on $t$. We can upper bound the number of cycles in Alg. \ref{AlgoMain} by the number of active vertices. Therefore, the total number of tests for the given adaptive $\F(t,s,\l)$-searching algorithm does not exceed $s\l(\log_2 t + F_{1}(s,\l) + F_{2}(s,\l) + 1)$.\quad $\square$
\begin{algorithm}[h]\label{AlgoMain}
 \KwData{set of vertices $V$ of  $H(V,E)\in \F(t,s,\l)$}
 \KwResult{set of edges of $H_{un}$}
initialization $E':=\emptyset$; $F: =\emptyset$; $S: =V$\;
 \While{$S\neq\emptyset$}{
  perform  Alg. \ref{AlgoSearchVertex}, and find  $v\not\in F$ \;
  $F: = F \sqcup v$\;
  perform  Alg. \ref{AlgoSearchEdges}, and find subset of edges $E'$\;
  perform  Alg. \ref{AlgoSearchQuery}, and find query $S$\;
  }
  set of edges $E'=E$\;
 \caption{Searching edges of the hidden hypergraph}
\end{algorithm}
\begin{algorithm}\label{AlgoSearchVertex}
 \KwData{query $S\subseteq V$ such that $Q(S) = 1$, and the set of found active vertices $F$}
 \KwResult{vertex $v\in V$, $v\not\in F$, and $\exists\, \e\in E$, $v\in \e$}
initialization $S': = S\setminus F$;  $S'' := S\setminus S'$\;
 \While{$|S'| > 1$}{
  split up  $S'$ into two subsets $S_1$ and $S_2$ of sizes $\lceil|S'|/2\rceil$ and $ \lfloor|S'|/2\rfloor$: $S' = S_1\sqcup S_2$\;
  carry out a query $S_1\sqcup S''$\;
  \eIf{$Q(S_1\sqcup S'') = 1$} 
  {
  $S': = S_1$\;
  }
  {
   $S':= S_2$\;
   $S'':=S''\sqcup S_1$\;
  }
  }
  vertex $\{v\}=S'$ satisfies the required conditions\;
 \caption{Searching another active vertex on the query}
\end{algorithm}
\begin{algorithm}\label{AlgoSearchEdges}
 \KwData{subset of active vertices $F\subset V$}
 \KwResult{subset of edges $E'\subset E$ consisting of vertices of $F$}
 initialization  $E': = \emptyset$\; 
 \For{$\forall S\subset F$: $1\le|S|\le\l$}{
  \eIf{$\nexists\, \e\in E':$ $\e\subset S$}{
  carry out query $S$\;  
  \eIf{Q(S) = 1
  }{
  \For{$\forall \e\in E':$ $S\subset \e$}{
  delete $\e$ from $E'$\;
  }
  add edge $\e = S$ to $E'$\;
  }
  {
  proceed to the next step of the loop\;
  }
   }{
   proceed to the next step of the loop\;
  }
  }
 
 \caption{Searching edges composed on found active vertices}
\end{algorithm}
\begin{algorithm}\label{AlgoSearchQuery}
 \KwData{subset of edges $E'\subset E$}
 \KwResult{or $S\subset V$ such that $Q(S) =1$, $\e\not\subset S$ for $\forall \e\in E'$, either $S=\emptyset$}
 initialization $A := \{v: \; v\in \e\in E'\}$; $B: = V\setminus A$; $S: = \emptyset$\;
 \For{$\forall C\subset V$: $B\subset C$ and $\nexists\, \e\in  E'$, $\e\subset C$ }{
  carry out query $C$\;
  \eIf{Q(C) = 1} 
  {
  $S := C$\;
  break ``for loop''\;
  }
  {
   proceed to the next step of the loop\;
  }  
  }
 \caption{Searching a query on found edges}
\end{algorithm}

\end{document}